\documentclass[aps,reprint,twocolumn,showpacs,floatfix]{revtex4-1}
\usepackage{amsmath,bm,epsfig}
\usepackage{latexsym}
\usepackage{amssymb}
\usepackage{color}
\usepackage{morefloats}
\usepackage{subfigure}
\usepackage{graphicx}
\usepackage{amsmath}
\usepackage{bm}
\usepackage{epsf,epsfig,graphics}
\usepackage{float}
\usepackage{makeidx}
\usepackage{graphicx}
\usepackage{natbib}
\usepackage{epsfig}
\usepackage{rotate}
\usepackage{float}
\usepackage{amssymb}
\usepackage{amsbsy}
\usepackage{overpic}
\usepackage{marvosym}
\usepackage{textcomp}
\usepackage{amsmath}
\usepackage{longtable}
\usepackage{graphicx}
\usepackage{dcolumn}
\usepackage{bm}

\newcommand\Nu{\text{Nu}}
\newcommand\Ra{\text{Ra}}
\newcommand\Rey{\text{Re}}
\newcommand\Pran{\text{Pr}}

\usepackage{color}

\newcommand{\blue}{\color{black} }

\newcommand{\bb}{\color{black} }

\begin{document}

\title{Heat Flux and Wall Shear Stress in Large Aspect-Ratio Turbulent Vertical Convection}
\author{Emily S.C. Ching}
\email{ching@phy.cuhk.edu.hk} \affiliation{Department of Physics,
The Chinese University of Hong Kong, Shatin, Hong Kong}
\date{\today}


\begin{abstract}
We present a theoretical analysis of large aspect-ratio turbulent
vertical convection that yields two relationships between heat flux and wall
shear stress, measured respectively by the Nusselt number ($\Nu$) and
shear Reynolds number ($\Rey_\tau$), in terms of the
 Rayleigh ($\Ra$) and Prandtl numbers ($\Pran$): $\Rey_\tau^2 \Nu = f(\Pran) \Pran^{-1} \Ra$ in
the high-$\Ra$ limit and $\Nu \approx C \Pran^{\varepsilon} \Rey_\tau$ with $\varepsilon=1/3$ for $\Pran \gg1$ and $\epsilon=1$ for $\Pran \ll 1$,
where $f(\Pran)$ is not a power law of $\Pran$ and $C$ is a
constant. These relationships imply $\Nu \approx [C^2f(\Pran)]^{1/3}
\Pran^{-(1-2\varepsilon)/3} \Ra^{1/3}$ and $\Rey_\tau \approx [f(\Pran)/C]^{1/3}
\Pran^{-(1+\varepsilon)/3} \Ra^{1/3}$ for high $\Ra$.

\end{abstract}


\maketitle

\begin{figure}[h!]
\vspace{-0.4cm}
    \centering
    \includegraphics[width=0.85\linewidth]{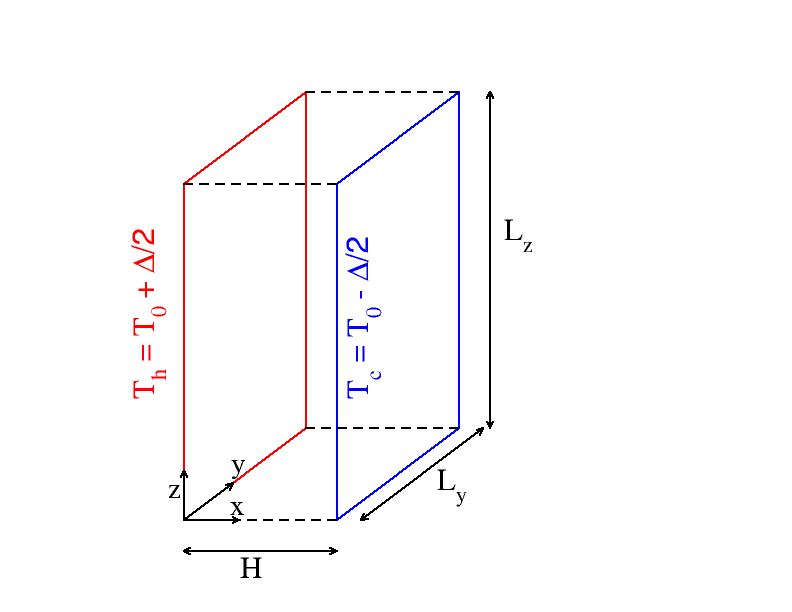}
     \vspace{-0.5cm}
    \caption{A schematic diagram of vertical convection. The left vertical wall is heated while the right vertical wall is cooled and the temperature difference is $\Delta=T_h-T_c$.}
    \label{Fig1}
   \vspace{-0.3cm}
\end{figure}

In astrophysical, geophysical and industrial fluid flows, fluid motions are often driven thermally by temperature differences. There are two common model systems for thermally driven flows: Rayleigh-B\'enard convection in a fluid heated from below and cooled from above~(see e.g. \cite{AGL2009,LX2010,CS2012,C2014}) and vertical convection in
a fluid between two vertical walls as shown in Fig.~\ref{Fig1}.  In these two systems, the direction of gravity makes a different angle to the boundaries that have a temperature difference. One important question in the study thermally driven fluid flows is how the heat transfer depends on the control parameters of the flow. Heat flux is commonly measured by the Nusselt number ($\Nu)$ and the control parameters include the Rayleigh number $\Ra \equiv \alpha g \Delta H^3/(\nu \kappa)$, which measures the strength of thermal forcing, and the Prandtl number of the fluid $\Pran \equiv \nu/\kappa$. Here, $H$ is the separation between the two boundaries with a temperature difference $\Delta$, $g$ the acceleration due to gravity, and $\alpha$, $\nu$, and $\kappa$ are the thermal expansion coefficient, kinematic viscosity and thermal diffusivity of the fluid, respectively. A scaling theory has been developed by Grossmann and Lohse~\cite{GL2000,GL2001,GL2002,GL2004}, which has successfully accounted for $\Nu(\Ra,\Pr)$ for a wide range of $\Ra$ and $\Pran$ in Rayleigh-B\'enard convection. In this Letter, we study vertical convection, which is much less studied than Rayleigh-B\'enard convection. Besides fundamental interest, vertical convection has many applications in engineering such as ventilation of buildings, thermal insulation in double-pane windows and cooling of electronic devices. It also plays a crucial role in ice-ocean interactions in which fluid motion is driven by temperature difference as well as the salinity difference between the melt water and the salty seawater~\cite{GGK2016,HVL2022}. For laminar vertical convection, analysis of the steady-state boundary-layer equations gives $\Nu\sim \Ra^{1/4}$~\cite{O1953,B1954,K1968,S2016} and the dependence of $\Nu$ on $\Pran$ in the low- and high-$\Pran$ limits~\cite{S2016}. When the flow becomes turbulent, fluctuations cannot be neglected and a theoretical understanding of the dependence of $\Nu$ on the control parameters is yet to be attained.

Turbulent vertical convection has been investigated experimentally and by direct numerical simulations (DNS). In most of these studies, $\Pran$ is kept fixed and a range of $\Ra$ is studied. The dependence of $\Nu$ on $\Ra$ has often been reported in the form of a power law: $\Nu \sim \Ra^\beta$. Experimental studies for a variety of fluids showed that $\beta$ changes from $1/4$, the value determined for laminar flow, to $1/3$ when the flow becomes turbulent~\cite{J1946,ME1969,CC1975,TN1988}. Results consistent with $\beta=1/3$ have been found in DNS in three dimensions with periodic boundary conditions in the spanwise ($y$) and streamwise ($z$) directions for $\Pran=0.709$ (air) and $\Ra$ between $10^5$ to $10^7$~\cite{VN1999,NCO2013} but other values of $\beta$ were reported in DNS at the same value of $\Pran$ such as $\beta=1/3.2$ for a similar range of $\Ra$~\cite{KH2012} and $\beta=0.31$ for a larger range of $\Ra$ up to $10^9$~\cite{NOLC2015}. It was pointed out that the value of $\beta$ depends on the range of $\Ra$ and $\Nu(\Ra)$ may not be a pure power law~\cite{NOLC2015}. The dependence on $\Pran$ has also been investigated for $1 \le \Pran \le 100$ and $10^6 \le \Ra \le 10^9$ and an effective power-law dependence of $\Nu$ on both $\Ra$ (with $\beta=0.321$) and $\Pran$ was reported~\cite{HNVL2022}. Effective power-law dependence $\Ra$ and $\Pran$ have also been reported for  the wall shear stress and the maximum mean vertical velocity of the convective flow~\cite{HNVL2022}. For DNS in two dimensions with adiabatic boundary condition in the horizontal boundaries, $\beta$ has been found to be closer to $1/4$ than $1/3$ for $\Pran=0.71$ and $6 \times 10^8 \le \Ra \le 10^{10}$~\cite{XQ1995,TSOP2007} but a recent study at $\Pran=10$ and $\Ra$ up to $10^{14}$ shows that there is a sharp transition from $\beta=1/4$ to $\beta=1/3$ when $\Ra \ge 5 \times 10^{10}$~\cite{WLVSL2021}. There have been different theoretical attempts to understand turbulent vertical convection. One approach is to identify relevant length, velocity and temperature scales in different flow regions and develop scaling functions of velocity and temperature in each region~\cite{GC1979,HH2005,BHH2007,W2019}. By matching the scaling functions of temperature in an assumed overlap region of two flow regions, expressions for $\Nu(\Ra)$ can be obtained and different results have been reported~\cite{GC1979,BHH2007}. Another study has tried to extend the scaling theory of Grossmann and Lohse~\cite{GL2000,GL2001,GL2002,GL2004} for Rayleigh-B\'enard convection to vertical convection but found that this approach is not feasible~\cite{NOLC2015}.

In this Letter, we present a theoretical analysis that yields two relationships between heat flux and wall
shear stress and their dependence on $\Ra$ and $\Pran$ in the high-$\Ra$ limit. We test the theoretical results for high $\Pran$ against the 
openly available DNS data for $1 \le \Pran \le 100$ and $10^6 \le \Ra \le 10^9$~\cite{HNVL2022} and find excellent agreement.

We consider a fluid confined between two vertical walls, 
with the left wall heated at a temperature $T_h$ 
and the right wall cooled at a temperature $T_c$ and the
temperature difference $\Delta$ is equal to $T_h-T_c$~(see Fig.~\ref{Fig1}). 
With the Oberbeck-Boussinesq approximation which neglects the variation of temperature in the fluid for all purposes except for the determination of the buoyancy force, the
governing equations are
\begin{eqnarray}
    \frac{\partial \mathbf{u} }{\partial t}+{\mathbf{u}}\cdot\nabla {\mathbf{u}}&=&- \frac{1}{\rho} \mathbf{\nabla} p + \nu\nabla^2 {\mathbf{u}}+\alpha g(T-T_0)\hat{z}
    \label{Meqn}
    \\
 \frac{\partial T}{\partial t}+{\mathbf{u}}\cdot\nabla T&=&\kappa\nabla^2T
    \label{Teqn}
    \\
    \nabla\cdot{\mathbf{u}}&=&0 \label{divfree}
\end{eqnarray}
where $\mathbf{u}(x,y,z,t)=(u,v,w)$ is the velocity, $p(x,y,z,t)$ the pressure,
$T(x,y,z,t)$ the temperature, $T_0$ the average temperature
of the two vertical plates and $\rho$ the density of the fluid at $T=T_0$. The coordinate system is shown in
Fig.~\ref{Fig1}  and $\hat{z}$ is a unit vector along
the vertical direction. The velocity field satisfies the no-slip boundary condition
at the two vertical plates.The flow quantities are Reynolds decomposed into sums of time
averages and fluctuations, e.g., $u(x,y,z,t)=U(x,y,z)+u'(x,y,z,t)$
and $T(x,y,z,t)-T_0=\Theta(x,y,z)+\theta'(x,y,z,t)$. 
We focus at the large aspect-ratio limit, namely  ${\blue L_z}/H \gg 1$ and
${\blue L_y}/H \gg 1$, where ${L_z}$ and ${L_y}$ are the height and width of the vertical walls.
In this limit, all the mean flow quantities depend on
$x$ only. Using the continuity equation Eq.~(\ref{divfree}) and
the no-slip boundary condition, we obtain $U=0$. Taking time
average of Eqs.~(\ref{Meqn}) and (\ref{Teqn}) leads to the mean
momentum balance and mean thermal energy balance equations~\cite{VN1999}
\begin{eqnarray}
    \frac{d}{dx} \langle u'{\blue w'}\rangle_t &=&\nu \frac{d^2}{dx^2} {\blue W} +\alpha g \Theta
   \label{MBE}\\
    \frac{d}{dx} \langle u'\theta'\rangle_t&=&\kappa \frac{d^2}{dx^2} \Theta
    \label{MTE}
\end{eqnarray}
where $\langle \cdots \rangle_t$ denotes an average over time. In
DNS where the computational domain is
finite, the same equations can be derived for
the mean quantities averaged over time as well as over $y$ and
$z$ if periodic boundary conditions are enforced in the $y$-
and $z$-directions~\cite{HNVL2022}. Due to the symmetry of
the problem, the mean velocity and temperature profiles ${\blue W}(x)$ and
$\Theta(x)$ are antisymmetric about $x=H/2$ thus one only has to
study Eqs.~(\ref{MBE}) and (\ref{MTE}) for $0 \le x \le H/2$. The
boundary conditions are
\begin{equation}
\begin{aligned}
    &{\blue W}(0)={\blue W}(H/2)=\Theta(H/2)=0;&\quad   &\Theta(0)=\Delta/2
    \label{BC}
\end{aligned}
\end{equation}

Integrating Eq.~(\ref{MTE}) with respect to $x$, one obtains
\begin{equation}
 \langle u'
\theta' \rangle_t - \kappa \frac{d\Theta}{dx} = -\kappa
\frac{d\Theta}{dx} \bigg|_{x=0} \label{MTEI}
\end{equation} showing that the mean horizontal heat flux $Q=\rho c \langle u' \theta' \rangle_t - k d\Theta/dx$ along the $x$
direction is independent of $x$~\cite{GC1979}. Here, $c$ and $k$ are the
specific heat capacity and thermal conductivity of the fluid,
respectively. $\Nu$ is defined as the ratio of actual heat flux to that when there were only thermal conduction, thus
\begin{equation}
\Nu \equiv \frac{Q}{k \Delta/H} = -\frac{d\Theta}{dx}
\bigg|_{x=0}\frac{H}{\Delta} = \frac{H}{2\delta_T} \label{Nu}
\end{equation}
where $\delta_T$ is the thermal boundary layer thickness defined by
$ \delta_T \equiv k\Delta/(2 Q)$. 
Integrating Eq.~(\ref{MBE}) with respect to $x$ gives~\cite{GC1979}
{\bb \begin{equation}
\langle u'w' \rangle_t = \nu \frac{dW}{dx} + \alpha g \int_0^{x} \Theta(x') dx'  -  \nu \frac{dW}{dx}\bigg|_{x=0} 
\label{MBEI}
\end{equation}}
The wall shear stress is given by $\tau_w = \rho \nu d W/dx |_{x=0}$  and is often measured by the dimensionless shear Reynolds
number $\Rey_\tau \equiv u_\tau H/\nu$ in terms of the friction velocity $u_\tau \equiv \sqrt{\nu d{\blue W}/dx|_{x=0}}$. {\bb If $W(x)$ and $\Theta(x)$ could be solved, then their gradients at $x=0$ or, equivalently, $\Rey_\tau$ and $\Nu$ would be obtained but Eqs.~(\ref{MBE}) and (\ref{MTE}) are not closed due to the presence of the second-order correlations $\langle u' w' \rangle_t$ and $\langle u' \theta' \rangle_t$. This is the well-known closure problem of turbulence in which there are more unknowns than equations. Our first step to tackle this problem is to evaluate Eq.~(\ref{MBEI}) at $x=x_0$, the location at which the magnitudes of the Reynolds shear stress and viscous stress are equal, i.e., $\nu {dW}/{dx}|_{x=x_0} = \langle u'w'\rangle_t (x_0)$.}
Near the wall, the viscous
stress {\blue $\rho \nu dW/dx$} dominates over the Reynolds shear stress $- \rho \langle u'{\blue w'} \rangle_t$, which is small and positive. As one
moves away from the wall, the viscous stress decreases while the Reynolds shear stress becomes negative and increases in magnitude.
Towards the centerline $x=H/2$, the viscous stress becomes negative and the magnitude of the Reynolds shear stress dominates over that 
of the viscous stress. The magnitudes of the two stresses are equal at $x=x_0$.
Evaluating Eq.~(\ref{MBEI}) at $x=x_0$ thus bypasses the difficulty of estimating $\langle u'w' \rangle_t$ and gives
\begin{equation}
\nu \frac{dW}{dx}\bigg|_{x=0} = \alpha g \int_0^{x_0} \Theta(x')
dx' \label{wallshearstress}
\end{equation}
Equation~(\ref{wallshearstress}) shows explicitly that
the wall shear stress $\tau_w$ is
generated by buoyancy and is equal to the buoyancy force per unit
area within the velocity boundary layer with $x \le x_0$. We define a
dimensionless temperature function $\Phi(\xi)$ of $\xi \equiv x/\delta_T$ by
\begin{equation}
\Theta({\bb x=\xi \delta_T}) \equiv \Delta \Phi(\xi) /2, \label{Phi}
\end{equation}
and rewrite Eq.~(\ref{wallshearstress}) to yield a universal
relation between $\Rey_\tau$ and $\Nu$:
\begin{equation}
\Rey_\tau^2 \Nu \Pran \Ra^{-1} = \frac{1}{4} \int_0^{\xi_0}
\Phi(\xi) d \xi \equiv I(\Ra,\Pran) \label{relation1}
\end{equation}
where $\xi_0=x_0/\delta_T$. Using Eqs.~(\ref{BC}) and (\ref{Nu}),
we obtain the boundary conditions for $\Phi$:
\begin{equation}
\Phi(0)=1, \ \  \Phi(\Nu)=0, \ \ \Phi'(0)=-1 \label{BCPhi}
\end{equation}

\begin{figure}[h!]
\vspace{-1.45cm}
    \centering
    \includegraphics[width=0.95\linewidth]{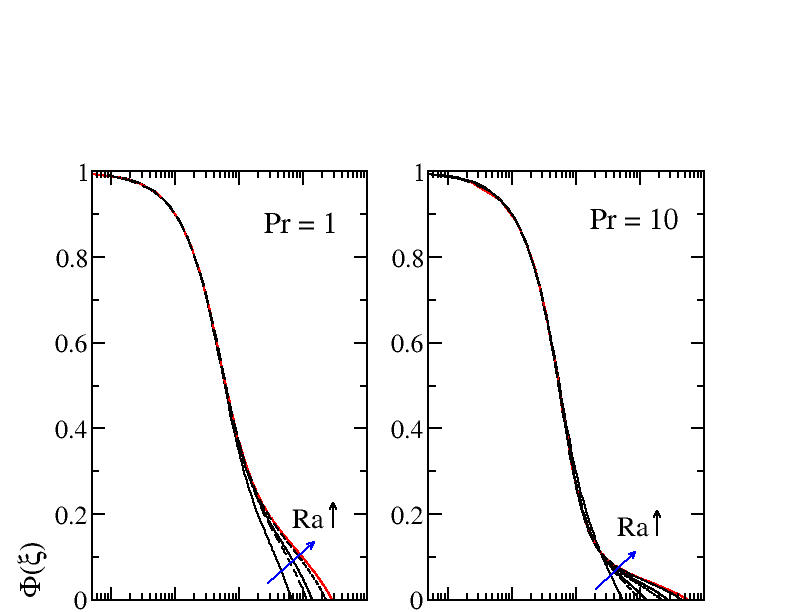}
    \includegraphics[width=0.95\linewidth]{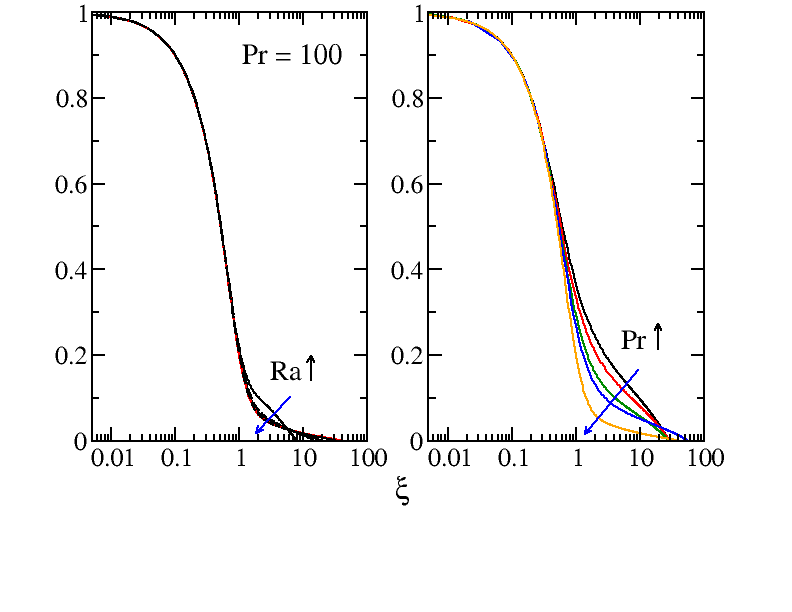}
    \vspace{-1.3cm}
    \caption{{\blue Plots of $\Phi(\xi)$ versus $x$ showing its dependence on $\Ra$ for $\Pran=1,10, 100$
    and its dependence on $\Pran$ at the largest $\Ra$ ($10^8$ for $\Pran=1, 2, 5$ and $10^9$ for $\Pran=10, 100$) using DNS data from Howland et al.~\cite{HNVL2022}. }}
    \label{Fig2}
\end{figure}

\begin{figure}[h!]
\vspace{-0.4cm}
    \centering
    \includegraphics[width=0.95\linewidth]{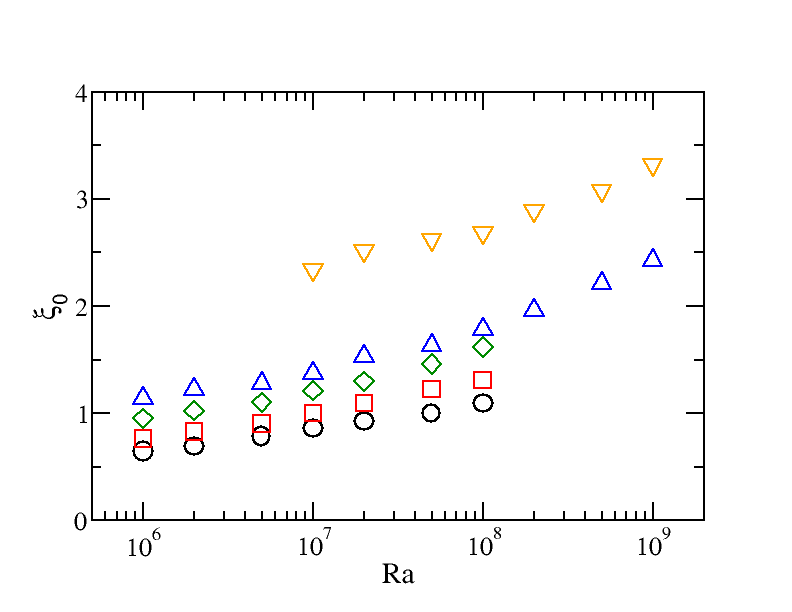}
    \vspace{-0.2cm}
    \caption{Dependence of $\xi_0$ on $\Ra$ for $\Pran=1$~(circles), $2$~(squares), $5$~(diamonds), $10$~(triangles) and $100$~(inverted triangles).}
    \label{Fig3}
\vspace{-0.6cm}
\end{figure}

We evaluate $\Phi(\xi)$ and $\xi_0$ using the DNS data from Howland et al.~\cite{HNVL2022,footnote} to study the $\Ra$- and $\Pran$-dependence of the integral $I$. 
As shown in Fig.~\ref{Fig2}, $\Phi(\xi)$ approaches an asymptotic form in the high-$\Ra$ limit for each $\Pran$ and the asymptotic form depends on $\Pran$.  In Fig.~\ref{Fig3}, it can be seen that $\xi_0$ increases slowly with $\Ra$ for each $\Pran$. Since $\Phi(\xi)$ decreases to zero for large $\xi$, these results suggest that the integral $I$ tends to a $\Ra$-independent function $f(\Pran)$ in the high-$\Ra$ limit. The available DNS data cover $1\le \Pran \le 100$ but we expect $\Phi(\xi)$ to approach an asymptotic form for general $\Pran$. For $\Pran < 1$, as the velocity boundary layer is nested within the thermal boundary layer, $\xi_0$ has to be less than $1$ and thus cannot increase with $\Ra$ asymptotically. It is expected that $\xi_0$ approaches a constant in the high-$\Ra$ limit for $\Pran <1$ and $I(\Ra,\Pran) \to f(\Pran)$ also for $\Pran<1$. Hence, we assume that
 \begin{equation}
\Rey_\tau^2 \Nu \Pran \Ra^{-1} = f(\Pran) \qquad \mbox{high-$\Ra$ limit} \label{relation1n}
\end{equation}

\begin{figure}[h!]
\vspace{-0.3cm}
    \centering
    \includegraphics[width=0.95\linewidth]{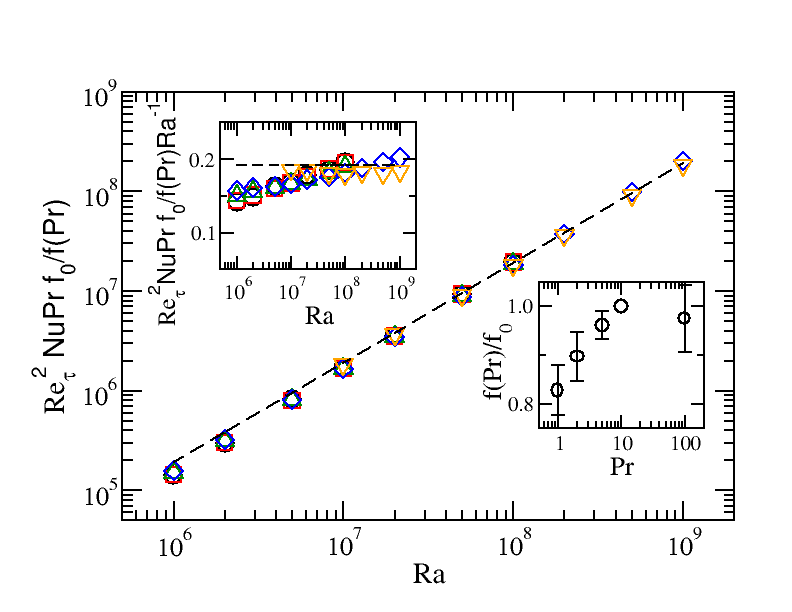}
    \caption{Dependence of $\Rey_\tau^2 \Nu \Pran f_0/f(\Pran)$ on $\Ra$ using the DNS data~\cite{HNVL2022} for $\Pran=1, 2, 5, 10, 100$ with same symbols as in Fig.~\ref{Fig3}. The dashed line is the best fit of the data points for $\Ra \ge 5 \times 10^7$ by the function $y=ax$ {\blue and the fitted value of $a$ gives $f_0=0.19$.} The inset on the left shows the compensated plots while the inset on the right shows $f(\Pran)/f_0$ versus $\Pran$.}    \label{Fig4}
\vspace{-0.3cm}
\end{figure}

We estimate the values of $f(\Pran)$ from the DNS
data~\cite{HNVL2022} as follows. Among the 38 sets of data, most data
were taken at $\Pran=10$. Thus we take $f_0 \equiv
f(\Pran=10)$ as a reference and estimate the values of
$f(\Pran)/f_0$ by the averages of the ratio of the data points
$\Rey_\tau^2 \Nu \Pran \Ra^{-1}$ for each of the other values of $\Pran$ to the data points at
 $\Pran=10$, taken at the 7 common values of $\Ra$.
{\blue The errors of the estimated $f(\Pran)/f_0$ are measured by the standard deviations.}
Equation~(\ref{relation1n}) implies that the data points of
$\Rey_\tau^2 \Nu$ for different values of $\Pran$, when multiplied
by $\Pran f_0/f(\Pran)$, would collapse into a single curve of
$f_0 \Ra$ for large $\Ra$. This is confirmed in Fig.~\ref{Fig4} and as shown in the inset of Fig.~\ref{Fig4},
$f(\Pran)/f_0$ cannot be approximated by a power law.

Using Eqs.~(\ref{Nu}) and (\ref{Phi}) we rewrite Eq.~(\ref{MTEI}) as
\begin{equation} \frac{\langle {\bb u' \theta'} \rangle_t}{\nu \Delta/H}
= \Nu \Pran^{-1} [1+ \Phi'(\xi)] \label{result3}
\end{equation}
Because of the boundary conditions, {\bb $u'$, $\theta'$ and $\partial u'/\partial x=-(\partial v'/\partial y+ \partial w'/\partial z)$} vanish at $x=0$.
As a result, $\langle u' \theta' \rangle_t$
and its first and second-order derivatives with respect to $x$
vanish at $x=0$ while $d^3 \langle u' \theta' \rangle_t/dx^3
|_{x=0}=  3 \langle {\partial^2 u'}/{\partial x^2} {\partial
\theta'}/{\partial x} \rangle_t |_{x=0}$. Taking the third-order
derivative of Eq.~(\ref{result3}) with respect to $\xi$ at $\xi=0$ gives
\begin{equation}
\frac{3 H^4}{{\blue 8} \nu \Delta} \Nu^{-3} \left \langle
\frac{\partial^2 u'}{\partial x^2}\frac{\partial \theta'}{\partial x}
\right \rangle_t \bigg|_{x=0}  = \Nu \Pran^{-1} \Phi^{(4)}(0)
\label{result4}
\end{equation}
{\bb Next, we make a closure estimate of $\langle ({\partial^2 u'}/{\partial x^2})({\partial
\theta'}/{\partial x}) \rangle_t |_{x=0}$. Physically we expect it to be related to the wall shear stress $\tau_w$ and the heat flux $Q$ and, therefore, depends on $u_\tau$ and $-d \Theta/dx |_{x=0}$. Near the wall, the molecular diffusivities are significant and we take the characteristic length scale to be $l_c=\nu/u_\tau$ for $\Pran \gg 1$ and $l_c=\kappa/u_\tau$ for $\Pran \ll 1$. Thus we let $\langle ({\partial^2 u'}/{\partial x^2})({\partial
\theta'}/{\partial x}) \rangle_t |_{x=0}=F(u_\tau, -d \Theta/dx |_{x=0}, l_c)$ and estimate the function $F$ by dimensional analysis to obtain \begin{equation}
\left \langle \frac{\partial^2 u'}{\partial x^2}\frac{\partial
\theta'}{\partial x} \right \rangle_t \bigg|_{x=0} \approx  c_0
\frac{u_\tau}{l_c^2} \frac{\Nu \Delta}{H} \label{approx}
\end{equation}
We have used Eq.~(\ref{Nu}) to write $-d\Theta/dx |_{x=0}=\Nu \Delta/H$.
Substituting Eq.~(\ref{approx}) into Eq.~(\ref{result4}), we obtain
\begin{equation}
 \Nu \approx C \Pran^{\varepsilon} \Rey_\tau, \qquad \varepsilon = \begin{cases} 1/3 & \Pran \gg 1 \\ 
 1 & \Pran \ll 1 
 \end{cases} 
 \label{relation2}
\end{equation} 
where $C = \{3 c_0/[8 \Phi^{(4)}(0)]\}^{1/3}$ is approximated as a constant, neglecting the possible weak $\Pran$-dependence of $\Phi^{(4)}(0)$.
Equation~(\ref{relation2}) for $\Pran \gg1$ agrees with the numerical result  $\Nu \sim \Pran^{1/3} \Rey_\tau$ found for $1\le \Pran \le 100$~\cite{HNVL2022}.  A relationship $\Nu \propto [\gamma(\Pran) \Pran]^{1/3}  \Rey_\tau$, where $\gamma(\Pran)$ is an undetermined function of $\Pran$, has been obtained for high $\Pran$ by assuming that the eddy diffusivity, defined by $-\langle u' \theta' \rangle_t/(\partial \Theta/\partial x)$, can be approximated by a cubic function of $x$, $\gamma(\Pran) u_\tau^3 x^3 /\nu^2$, throughout the thermal boundary layer~\cite{RF1980} but the cubic-function approximation is valid only for a very small region close to the wall and does not hold for the whole thermal boundary layer. 

Solving Eqs.~(\ref{relation1n}) and (\ref{relation2}), we obtain 
{\bb\begin{eqnarray}
\Nu &\approx& [C^2 f(\Pran)]^{1/3} \Pran^{-(1-2\varepsilon)/3} \Ra^{1/3} \label{NuhighRa}\\
\Rey_\tau &\approx& [f(\Pran)/C]^{1/3} \Pran^{-(1+\varepsilon)/3} \Ra^{1/3} \label{RetauhighRa} 
\end{eqnarray}
in the high-$\Ra$ limit.}
These theoretical results imply that data points of $\Nu$ and $\Rey_\tau$ taken at different values of $\Pran$ can be collapsed into single curves of $\Ra^{1/3}$-dependence for large 
$\Ra$ when multiplied by appropriate factors of $f(\Pran)$ and  $\Pran$.  
As shown in Fig.~\ref{Fig5}, the theoretical predictions {\bb for high $\Pran$} are in excellent agreement with the available DNS data for $1 \le \Pran \le 100$~\cite{HNVL2022}. Data for low $\Pran$ and high $\Ra$ are not yet available.

\begin{figure}[th!]
\vspace{-0.2cm}
    \centering
    \includegraphics[width=0.95\linewidth]{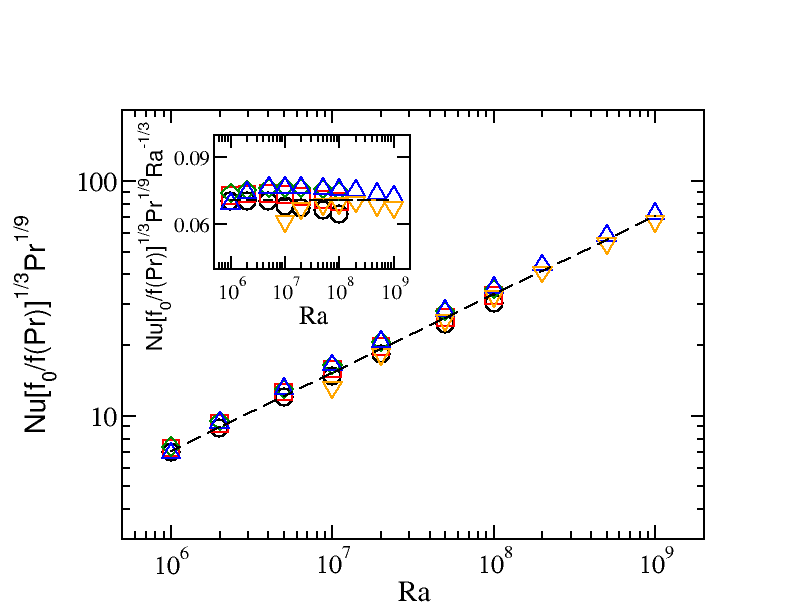}
    \includegraphics[width=0.95\linewidth]{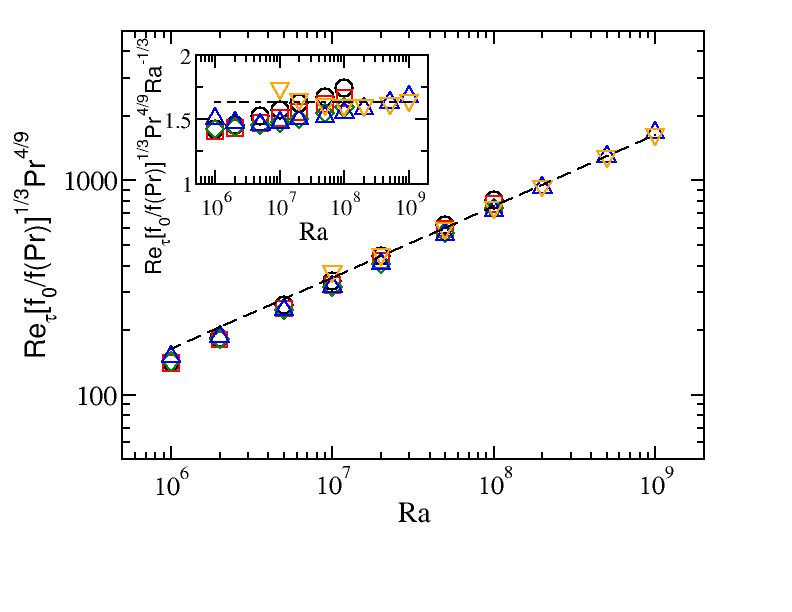}
    \vspace{-0.4cm}
    \caption{Dependence of ${\blue \Nu [f_0/f(\Pran)]^{1/3} \Pran^{1/9}}$ (top) and ${\blue \Rey_\tau [f_0/f(\Pran)]^{1/3} \Pran^{4/9}}$ (bottom) on $\Ra$ using the DNS data~\cite{HNVL2022} for $\Pran=1, 2, 5, 10, 100$ with same symbols as in Fig.~\ref{Fig3}.  The dashed lines are the best fits of the theoretical prediction $y \propto x^{1/3}$ for data points taken at $\Ra \ge 5 \times 10^7$ and the fitted values of the proportionality constants in the two fits give $f_0=0.19$ and $C=0.043$. The insets show the compensated plots.}
    \label{Fig5}
    \vspace{-0.5cm}
\end{figure}

In summary, we have obtained theoretical results for the dependence of $\Nu$ and $\Rey_\tau$ on $\Ra$ and $\Pran$, answering the question of how heat flux and wall shear stress depend on the control parameters for large aspect-ratio turbulent vertical convection in the high-$\Ra$ limit. Such a question is challenging because of the underlying closure problem of turbulence in which Eqs.~(\ref{MBE}) and (\ref{MTE}) for the mean quantities, $W(x)$ and $\Theta(x)$, contain additional unknowns of the second-order correlations, $\langle u'w' \rangle_t$ and $\langle u' \theta'\rangle_t$. Our theoretical analysis purposefully bypasses the difficulty of directly estimating $\langle u' w' \rangle_t$, assumes the integral $I(\Ra,\Pran)$ approaching a $\Ra$-independent function in the high-$\Ra$ limit and makes a minimal closure estimate of the third-order derivative of $ \langle u' \theta' \rangle_t$ at $x=0$ instead of the whole function. For finite $\Ra$, the additional $\Ra$-dependence of $I(\Ra,\Pran)$~[see Eq.~(\ref{relation1})], which is expected not in the form of a power law, would modify the $\Ra^{1/3}$-dependence of $\Nu$ and $\Rey_\tau$. This could explain the variations of the effective power-law exponent $\beta$ for $\Nu(\Ra)$ observed in different ranges of $\Ra$ in DNS~\cite{VN1999,KH2012,NCO2013,NOLC2015,HNVL2022}. The present work studies the limit of large aspect ratios but our theoretical result of $\Nu \sim \Ra^{1/3}$ in the high-$\Ra$ limit is also in agreement with the DNS result for a two-dimensional cell with unit aspect ratio in the turbulent regime~\cite{WLVSL2021}.

\acknowledgments
The author thanks Christopher J. Howland for providing the DNS data and Detlef Lohse and Olga Shishkina for discussions. She also acknowledges support from the Hong Kong Research Grants Council (Grant No. CUHK 14302419).


\begin{thebibliography}{35}
\bibitem{footnote}%
 Additional DNS data on the profiles of $\langle u' w' \rangle_t$ have been provided by Christopher J. Howland.
\end{thebibliography}
\end{document}